\documentclass[aps,twocolumn,showpacs]{revtex4}
\usepackage[dvips]{graphics}

\usepackage{color}
\definecolor{gold}{rgb}{0.85,0.66,0}
\definecolor{dblue}{rgb}{0,0,0.8}

\begin{document}

\title{{\textcolor{gold}{Multi-terminal quantum transport through a 
single benzene molecule: Evidence of a Molecular Transistor}}}

\author{{\textcolor{dblue}{Santanu K. Maiti}}$^{1,2}$}

\affiliation{$^1$Theoretical Condensed Matter Physics Division, 
Saha Institute of Nuclear Physics, 1/AF, Bidhannagar, Kolkata-700 064, 
India \\
$^2$Department of Physics, Narasinha Dutt College, 129 Belilious Road, 
Howrah-711 101, India} 

\begin{abstract}
We explore multi-terminal quantum transport through a benzene molecule
threaded by an Aharonov-Bohm flux $\phi$. A simple tight-binding model is 
used to describe the system and all the calculations are done based on 
the Green's function formalism. With a brief description of two-terminal
quantum transport, we present a detailed study of three-terminal
transport properties through the benzene molecule to reveal the actual
mechanism of electron transport. Here we numerically compute the
multi-terminal conductances, reflection probabilities and current-voltage 
characteristics in the aspects of molecular coupling strength and 
magnetic flux $\phi$. Most significantly we observe that, the molecular
system where the benzene molecule is attached to three terminals can
be operated as a transistor, and we call it a molecular transistor.
This aspect can be utilized in designing nano-electronic circuits and 
our investigation may provide a basic framework to study electron 
transport in any complicated multi-terminal quantum system. 
\end{abstract}


\maketitle

\section{Introduction}

The rapid development in nanofabrication techniques have enabled us to
measure current through molecular wires, even through a single isolated
molecule attached to two measuring electrodes. Electronic transport
through molecular systems have attracted much more attention since these 
are the fundamental building blocks for future generation of electronic 
devices. In $1974$, Aviram and Ratner~\cite{aviram} first studied 
theoretically the electron transport through a molecular bridge. In their 
work they have calculated two-terminal conductance based on the Landauer 
conductance formula~\cite{land}. Following this pioneering work, several
experiments have been done through different molecules placed between two 
electrodes with few nanometer separation. Though, to date a lot of 
theoretical~\cite{mag,lau,baer1,baer2,baer3,tagami,gold} as well as 
experimental works~\cite{reed1,reed2,tali,fish} on two-terminal 
electron transport have been done addressing several important issues, 
but a very few works are available on multi-terminal quantum
systems~\cite{xu1,zhao,ember,let,zhong,lin,pic,sumit} and still 
it is an open subject to us. 
B\"{u}ttiker~\cite{butt1} first addressed theoretically the electron 
transport in multi-terminal quantum systems following the theory of 
Landauer two-terminal conductance formula. Full quantum mechanical 
approach is needed to study electron transport in molecular systems. The 
transport properties are characterized by several significant factors like 
as the quantization of energy levels, quantum interference of electronic 
waves associated with the geometry of the bridging system adopts within 
the junction and other different parameters of the Hamiltonian that are 
used to describe a complete system. 
 
Several {\em ab initio} methods are used to study electron 
transport~\cite{yal,ven,xue,tay,der,dam} through molecular systems. At 
the same time, tight-binding model has extensively been studied in the 
literature and it has also been extended to DFT transport
calculations~\cite{elst}. The study of static density functional theory 
(DFT)~\cite{kohn1,kohn2} within the local-density approximation (LDA) to 
investigate the electron transport through nanoscale conductors, like 
atomic-scale point contacts, has met with great success. But, when this 
similar theory applies to molecular junctions, theoretical conductances 
achieve larger values compared to the experimental predictions and these 
quantitative discrepancies need detailed study in this particular field. 
In a recent work, Sai {\em et al.}~\cite{sai} have predicted a correction 
to the conductance using the time-dependent current-density functional 
theory since the dynamical effects give significant contribution to the 
electron transport, and demonstrated some key results with specific 
examples. Quite similar dynamical effects have also been reported in some 
other recent papers~\cite{bush,ven1}, where authors have abandoned the 
infinite reservoirs, as originally introduced by Landauer, and considered 
two large but finite oppositely charged electrodes connected by a 
nanojunction. In the present paper, we reproduce an analytic approach 
based on a simple tight-binding model to characterize the electron 
transport properties through a benzene molecule placed between the 
macroscopic contacts. A simple parametric approach~\cite{muj1,muj2,san1,
sam,san2,hjo,walc1,walc2} is presented for the calculations and it is 
motivated by the fact that the {\em ab initio} theories are 
computationally much more expensive, while the model calculations 
by using the tight-binding formulation are computationally very 
cheap and also provide a worth insight to the problem.

The aim of this work is to describe multi-terminal electron transport 
through a single benzene molecule threaded by an Aharonov-Bohm (AB) 
flux $\phi$. 
We do exact numerical calculation based on the Green's function formalism 
to reveal the actual mechanism of electron transport. With a very brief 
description of electron transport in two-terminal system, we express 
elaborately the three-terminal transport properties of the benzene 
molecule. Here we numerically compute the conductances, reflection 
probabilities and current-voltage characteristics as functions of the 
molecule-to-lead coupling strength and threaded magnetic flux $\phi$. 
Most interestingly we predict that the benzene molecule attached to 
three leads can be used as a transistor, and we can call it as 
a molecular transistor. These three leads are quite analogous to 
emitter, base and collector as defined in conventional transistor. 

The scheme of the paper is as follow. With the brief introduction 
(Section I), in Section II, we describe the model and the theoretical 
formulations for our calculations. Section III presents the significant 
results, and finally, we conclude our results in Section IV.

\section{Model and the synopsis of the theoretical background}

\subsection{Two-terminal molecular system}

\begin{figure}[ht]
{\centering \resizebox*{6cm}{1.85cm}{\includegraphics{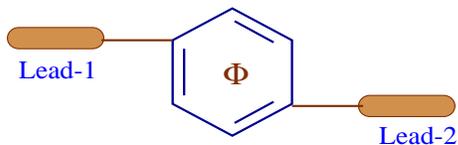}}\par}
\caption{(Color online). Two-terminal quantum system where a benzene 
molecule, threaded by a magnetic flux $\phi$, is attached symmetrically
to two semi-infinite one-dimensional metallic leads, viz, lead-$1$ and 
lead-$2$.}
\label{benzenetwo}
\end{figure}

Let us start by referring to Fig.~\ref{benzenetwo}, where a benzene 
molecule, threaded by a magnetic flux $\phi$, is attached symmetrically 
to two semi-infinite one-dimensional ($1$D) metallic leads, namely,
lead-$1$ and lead-$2$. 

To calculate two-terminal conductance ($g$) of the benzene molecule, 
we use the Landauer conductance formula~\cite{datta,marc}. At much low 
temperatures and bias voltage it can be expressed as,
\begin{equation}
g=\frac{2e^2}{h} T
\label{equ1}
\end{equation}
where, $T$ gives the transmission probability of an electron across the 
molecule and it is related to the reflection probability $R$ by the 
expression $R=1-T$. In terms of the Green's function of the molecule and 
its coupling to the leads, the transmission probability can be written 
in the form~\cite{datta,marc},
\begin{equation}
T={\mbox{Tr}} \left[\Gamma_1 G_{M}^r \Gamma_2 G_{M}^a\right]
\label{equ2}
\end{equation}
where, $\Gamma_1$ and $\Gamma_2$ describe the coupling of the molecule 
to the lead-$1$ and lead-$2$, respectively. Here, $G_{M}^r$ and $G_{M}^a$ 
are the retarded and advanced Green's functions, respectively, of the 
molecule including the effects of the leads. Now, for the full system 
i.e., the molecule and two leads, the Green's function is expressed 
as,
\begin{equation}
G=\left(E-H\right)^{-1}
\label{equ3}
\end{equation}
where, $E$ is the energy of the injecting electron. Evaluation of this 
Green's function needs the inversion of an infinite matrix, which is 
really a difficult task, since the full system consists of the finite 
size molecule and the two semi-infinite $1$D leads. However, the full 
system can be partitioned into sub-matrices corresponding to the 
individual sub-systems and the Green's function for the molecule can 
be effectively written in the form~\cite{marc,datta},
\begin{equation}
G_M=\left(E-H_M-\Sigma_1-\Sigma_2 \right)^{-1}
\label{equ4}
\end{equation}
where, $H_M$ corresponds to the Hamiltonian of the benzene molecule.
Within the non-interacting picture, the Hamiltonian of the molecule
can be expressed like,
\begin{equation}
H_M = \sum_i \epsilon_i c_i^{\dagger} c_i + \sum_{<ij>} t 
\left(c_i^{\dagger} c_j e^{i\theta}+ c_j^{\dagger} c_i e^{-i\theta}\right)
\label{equ5}
\end{equation}
Here, $\epsilon_i$ and $t$ correspond to the site energy and 
nearest-neighbor hopping strength, respectively. $c_i^{\dagger}$ ($c_i$) 
is the creation (annihilation) operator of an electron at the site $i$
and $\theta=\pi \phi/3 \phi_0$ is the phase factor due to the flux $\phi$ 
enclosed by the molecular ring. A similar kind of tight-binding 
Hamiltonian is also used, except the phase factor $\theta$, to describe 
the leads where the Hamiltonian is parametrized by constant on-site 
potential $\epsilon^{\prime}$ and nearest-neighbor hopping integral 
$t^{\prime}$. The hopping integral between the lead-$1$ and the
molecule is $\tau_1$, while it is $\tau_2$ between the molecule and 
the lead-$2$. In Eq.~(\ref{equ4}), $\Sigma_1$ and $\Sigma_2$ are the 
self-energies due to the coupling of the molecule to the lead-$1$ and 
lead-$2$, respectively, where all the information of the coupling are 
included into these self-energies.

The current passing through the molecule can be regarded as a single
electron scattering process between the two reservoirs of charge
carriers. The current-voltage relationship can be obtained from the
expression~\cite{marc,datta},
\begin{equation}
I(V)=\frac{e}{\pi \hbar}\int \limits_{-\infty}^{\infty} 
\left(f_1-f_2\right) T(E)~ dE
\label{equ6}
\end{equation}
where, $f_{1(2)}=f\left(E-\mu_{1(2)}\right)$ gives the Fermi distribution
function with the electrochemical potential $\mu_{1(2)}=E_F\pm eV/2$.
$E_F$ is the equilibrium Fermi energy. Usually, the electric field inside 
the molecule, especially for small molecules, seems to have a minimal 
effect on the $g$-$E$ characteristics. Thus it introduces a very little 
error if we assume that the entire voltage is dropped across the 
molecule-lead interfaces. On the other hand, for larger molecules
and higher bias voltage, the electric field inside the molecule may 
play a more significant role depending on the size and structure of 
the molecule~\cite{tian}, though the effect becomes quite small.

\subsection{Three-terminal molecular system}

\begin{figure}[ht]
{\centering\resizebox*{3.75cm}{6cm}{\includegraphics{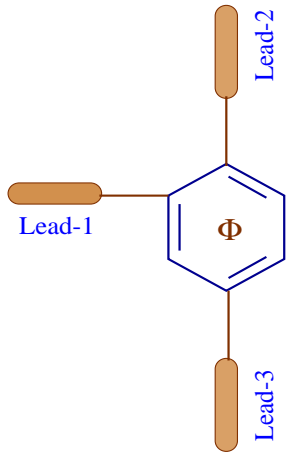}}\par}
\caption{(Color online). Three-terminal quantum system where a benzene 
molecule, threaded by a magnetic flux $\phi$, is attached asymmetrically
to three semi-infinite $1$D metallic leads, namely, lead-$1$, lead-$2$ 
and lead-$3$.}
\label{benzenethree}
\end{figure}

The schematic view of a three-terminal quantum system is presented in
Fig.~\ref{benzenethree}, where a benzene molecule is attached to three
semi-infinite leads, viz, lead-$1$, lead-$2$ and lead-$3$. These leads
are coupled to the molecule asymmetrically i.e., the path differences
between them are not identical to each other and they (three leads) 
are quite analogous
to the emitter, base and collector as defined in traditional macroscopic
transistor. The actual scheme of connections with the batteries for the 
operation
\begin{figure}[ht]
{\centering\resizebox*{5.5cm}{7.5cm}{\includegraphics{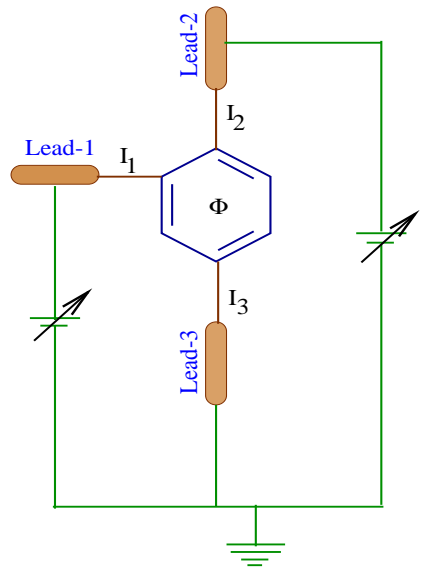}}\par}
\caption{(Color online). The scheme of connections with the batteries for
the operation of the benzene molecule as a transistor. The voltages in the 
lead-$1$ and lead-$2$ are applied with respect to the lead-$3$.}
\label{circuit}
\end{figure}
of the molecule as a transistor is depicted in Fig.~\ref{circuit}, where
the voltages in the lead-$1$ and lead-$2$ are applied with respect to
the lead-$3$.

In order to calculate the conductance in this three-terminal quantum 
system, we use B\"{u}ttiker formalism, an elegant and simple way to 
study electron transport through multi-terminal mesoscopic systems.
In this formalism we treat all the leads (current and voltage leads) on
an equal footing and extend the two-terminal linear response formula to 
get the conductance between the terminals, indexed by $p$ and $q$, in 
the form~\cite{datta}, 
\begin{equation}
g_{pq}=\frac{2e^2}{h} T_{pq}
\label{equ7}
\end{equation}
where, $T_{pq}$ gives the transmission probability of an electron 
from the lead-$p$ to lead-$q$. Here, the reflection probabilities 
are related to the transmission probabilities by the equation 
$R_{pp}+\sum_{q(\ne p)}T_{qp}=1$, which is obtained from the condition
of current conservation~\cite{xu}. Now, similar to Eq.~(\ref{equ2}), the
transmission probability $T_{pq}$ can be expressed in terms of the 
molecule-lead coupling matrices and the effective Green's function of 
the molecule as~\cite{datta},
\begin{equation}
T_{pq}={\mbox{Tr}} \left[\Gamma_p G_{M}^r \Gamma_q G_{M}^a\right]
\label{equ8}
\end{equation}
In the presence of multi leads, the effective Green's function of the
molecule becomes (extension of Eq.~(\ref{equ4}))~\cite{datta},
\begin{equation}
G_M=\left(E-H_M-\sum_p \Sigma_p \right)^{-1}
\label{equ9}
\end{equation}
where, $\Sigma_p$ is the self-energy due to the coupling of the molecule
to the lead-$p$ and the sum over $p$ runs from $1$ to $3$. $H_M$ is the
molecular Hamiltonian (see Eq.~(\ref{equ5})).

Finally, as an extension of two-terminal devices, we can write the current
$I_p$ for the lead-$p$ in the form~\cite{datta},
\begin{equation}
I_p(V)=\frac{e}{\pi \hbar} \sum_q \int \limits_{-\infty}^{\infty} 
T_{pq}(E) \left[f_p(E)-f_q(E) \right] dE
\label{equ10}
\end{equation}
In this presentation, all the results are computed only at absolute zero
temperature. These results are also valid even for some finite (low)
temperatures, since the broadening of the energy levels of the benzene
molecule due to its coupling to the leads becomes much larger 
than that of the thermal broadening~\cite{datta}. On the other hand, 
at high temperature limit, all these features completely disappear. 
This is due to the fact that the phase coherence length decreases 
significantly with the rise of temperature where the contribution 
comes mainly from the scattering on phonons, and therefore, the 
quantum interference effect vanishes. For the sake of simplicity, 
we take the unit $c=e=h=1$ in our present calculations

\section{Results and discussion}

To illustrate the results, let us begin our discussion by mentioning the 
values of the different parameters used for the numerical calculations. 
In the benzene molecule, the on-site energy $\epsilon_i$ is fixed to $0$ 
for all the sites $i$ and the nearest-neighbor hopping strength $t$ is 
set to $3$. While, for the side-attached leads the on-site energy 
($\epsilon^{\prime}$) and the nearest-neighbor hopping strength 
($t^{\prime}$) are chosen as $0$ and $4$, respectively. The Fermi energy 
$E_F$ is taken as $0$. Throughout the study, we narrate our results for 
the two limiting cases depending on the strength of the coupling of the 
molecule to the leads. Case I: $\tau_{1(2,3)} << t$. It is the so-called 
weak-coupling limit. For this regime we choose $\tau_1=\tau_2=\tau_3=0.5$. 
Case II: $\tau_{1(2,3)} \sim t$. This is the so-called strong-coupling 
limit. In this particular limit, we set the values of the parameters as 
$\tau_1=\tau_2=\tau_3=2.5$. 

In the forthcoming sub-sections we will describe the characteristic 
properties of electron transport both for the two- and three-terminal 
molecular systems and our exact study may give some significant insight 
to the electron transport in any multi-terminal quantum system.

\subsection{Two-terminal molecular system}

\subsubsection{Conductance-energy characteristics}

\begin{figure}[ht]
{\centering \resizebox*{8cm}{7cm}{\includegraphics{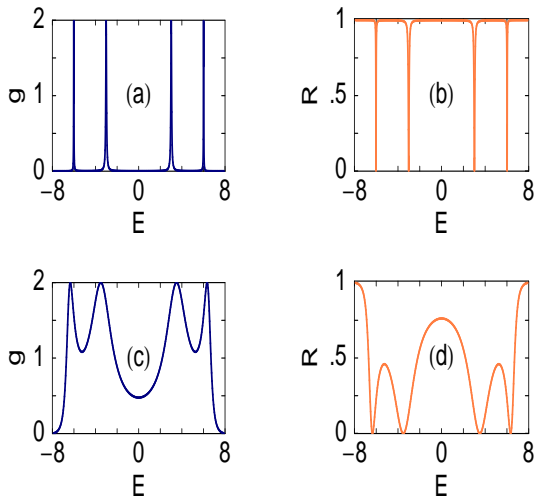}}\par}
\caption{(Color online). Two-terminal conductance $g$ and reflection 
probability $R$ as a function of the energy $E$ of the benzene molecule 
with $\phi=0$. (a) and (b) represent the results for the weak-coupling 
limit, while, (c) and (d) correspond to the same for the strong-coupling 
limit.}
\label{twocond}
\end{figure}

As illustrative examples, in Fig.~\ref{twocond} we present the variation 
of two-terminal conductance $g$ and reflection probability $R$ as
a function of the injecting electron energy $E$.
The results for the weak-coupling limit are shown in (a) and (b), while
(c) and (d), correspond to the variation for the limit of strong molecular 
coupling. The flux $\phi$ is fixed at $0$. In the limit of weak-coupling,
conductance exhibits sharp resonant peaks (Fig.~\ref{twocond}(a)) for some 
particular energies, while it vanishes almost for all other energies. It
emphasizes that a fine tuning in the energy scale is necessary to get 
electron conduction across the molecule. At the resonances, the 
conductance reaches the value $2$, and therefore, the transmission 
probability $T$ goes to unity, since the relation $g=2T$ is satisfied 
from the Landauer conductance formula (see Eq.~(\ref{equ1}) with $e=h=1$ 
in our present description). All these resonant peaks are associated with 
the energy eigenvalues of the benzene molecule, and hence, we can
predict that the conductance spectrum manifests itself the electronic 
structure of the molecule. Following this conductance spectrum, the 
variation of the reflection probability ($R$) (Fig.~\ref{twocond}(b)) 
can be clearly explained. It shows sharp dips ($R=0$) for some fixed 
energies where the conductance gets the value $2$. At these resonant 
energies $T=1$ which provides $R=0$, since for the two-terminal quantum 
system $R$ is related to the transmission probability $T$ by the equation 
$R=1-T$. For all other energies $R$ becomes $1$, which reveals that 
for these cases no electron conduction takes place through the molecule. 
The behavior of electron transport becomes quite interesting as long as 
the molecular coupling is increased. In the strong-coupling limit, all the
resonant peaks get substantial widths (Fig.~\ref{twocond}(c)) compared 
to the weak-coupling limit. The contribution to the broadening of the 
resonant peaks appears from the imaginary parts of the self-energies 
$\Sigma_1$ and $\Sigma_2$, respectively~\cite{datta}. From the conductance 
spectrum it is observed that the electron conduction through the molecular
bridge takes place almost for all energies, and therefore, in the strong 
molecular coupling, fine tuning in the energy scale is not required to get 
electron conduction across the molecule. A similar effect of the molecular 
coupling is also observed in the $R$-$E$ spectrum (Fig.~\ref{twocond}(d)). 
It is noticed that almost for the entire energy range the reflection
probability does not reach to unity anymore, which allows electron 
transmission, and, specifically for the four typical energies, it ($R$) 
drops exactly to zero which indicates ballistic transmission through the 
molecular wire.

\subsubsection{Effects of magnetic flux $\phi$}

\begin{figure}[ht]
{\centering \resizebox*{7.5cm}{4.65cm}{\includegraphics{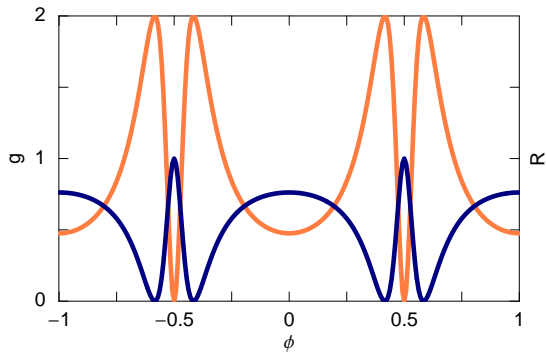}}\par}
\caption{(Color online). Two-terminal conductance $g$ (reddish yellow) and 
reflection probability $R$ (blue) in the limit of strong-coupling as a 
function of $\phi$ of the benzene molecule with $E=0$.}
\label{twocondphi}
\end{figure}

To visualize the effects of magnetic flux $\phi$ on electron transport, in
Fig.~\ref{twocondphi} we display the variation of two-terminal conductance
(reddish yellow line) and reflection probability (blue line) as a function 
of $\phi$. The results are computed for the typical energy $E=0$ in the
limit of strong molecular coupling. Both the conductance and reflection
probability vary periodically with $\phi$ showing $\phi_0$ ($=1$, in our
chosen unit) flux-quantum periodicity. In the presence of magnetic flux,
the $g$-$E$ spectrum gets modified significantly due to the additional phase
shift among the electronic waves traversing through different arms of the
molecular ring and one can control the electron transmission through the
molecular wire in a meaningful way. Most significantly we see that, for
the typical flux $\phi=\phi_0/2$, the conductance exactly vanishes, and
therefore $R$ becomes $1$, which reveals zero transmission. This feature
can be implemented as follow. The probability amplitude of getting an 
electron from the lead-$1$ to lead-$2$ across the molecular ring depends 
on the quantum interference of the electronic waves passing through
the upper and lower arms of the ring. For the symmetrically connected
molecule (lengths of the upper and lower arms of the molecular ring 
are identical to each other) which is threaded by a magnetic flux $\phi$, 
the probability amplitude of getting an electron across the molecule 
becomes exactly zero ($T=0$) for the typical flux, $\phi=\phi_0/2$. This 
is due to the result of the quantum interference among the two waves in 
the two arms of the molecular ring, which can be shown through few simple 
mathematical steps. Thus, quantum interference effect plays a crucial 
role in the study of electron transport in a molecular bridge system 
which can be controlled by changing the lead-molecule interface geometry 
as well as by changing the AB flux $\phi$ passing through the molecule.

\subsubsection{Current-voltage characteristics}

\begin{figure}[ht]
{\centering \resizebox*{7.5cm}{9.9cm}{\includegraphics{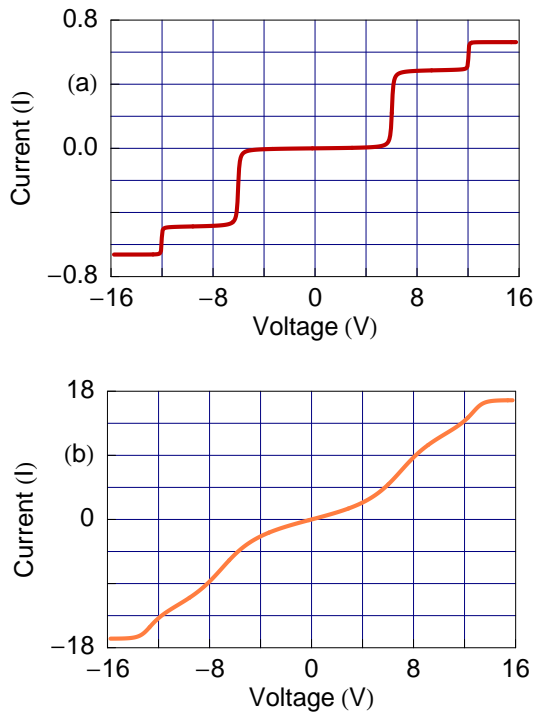}}\par}
\caption{(Color online). $I$-$V$ spectra of the benzene molecule with 
$\phi=0$. (a) and (b) correspond to the results for the weak- and 
strong-coupling limits, respectively.}
\label{twocurr}
\end{figure}

All the basic features of electron transfer through the molecule
become much more clearly visible by investigating the current-voltage 
characteristics. The current $I$ is determined by the integration
procedure of the transmission function $T$ (see Eq.~(\ref{equ6})), 
where the function $T$ varies exactly similar to the conductance 
spectra, differ only in magnitude by the factor $2$, since the 
equation $g=2T$ is satisfied from the Landauer conductance formula 
(Eq.~(\ref{equ1})). As representative examples, in Fig.~\ref{twocurr}
we display the variation of current with the bias voltage $V$ for the
benzene molecule considering $\phi=0$. The result for the weak-coupling
limit is shown in (a), while for the case of strong-coupling it is
presented in (b). In the weak-coupling case, the current shows
staircase like structure with sharp steps as a function of the 
applied bias voltage $V$. This is due to the presence of fine resonant
peaks in the conductance spectrum (Fig.~\ref{twocond}(a)), as the
current is computed from the integration procedure of the transmission
function $T$. The electrochemical potentials in the leads cross one
of the molecular energy levels as we increase the bias voltage,
and accordingly, it provides a jump in the current-voltage spectrum.
Addition to this, it is also important to note that the non-zero value 
of the current appears beyond a finite value of $V$, the so-called 
threshold voltage $V_{th}$. The behavior of $I$-$V$ characteristics 
changes significantly in the limit of strong molecular coupling
(Fig.~\ref{twocurr}(b)).
The step-like feature almost disappears and the current varies quite
continuously with the bias voltage $V$. Not only that, it ($I$) also
achieves very large current amplitude compared to the weak-coupling 
case. This enhanced current amplitude can be noticed clearly by observing 
the area under the $g$-$E$ curve presented in Fig.~\ref{twocond}(c). Here 
the non-zero value of the current is observed for very small value of the 
bias voltage $V$ i.e., $V_{th}\rightarrow 0$. Finally, from these $I$-$V$ 
spectra (Figs.~\ref{twocurr}(a) and (b)) it can be manifested that the
molecule-to-lead coupling strength has a significant influence in the 
determination of the current amplitude as well as the threshold bias 
voltage $V_{th}$, which may provide several key features in designing 
molecular electronic devices.

\subsection{Three-terminal molecular system}

Following the above brief description of electron transport in the 
two-terminal molecular system, now we focus our results in detail 
for the three-terminal molecular system, and here we will show how 
such a simple system can be utilized as an electronic transistor.

\subsubsection{Conductance-energy characteristics}

In the three-terminal molecular system, several anomalous features
are observed in the conductance-energy spectra as well as in the 
variation of reflection probability with the energy $E$.
\begin{figure}[ht]
{\centering \resizebox*{8cm}{10.5cm}
{\includegraphics{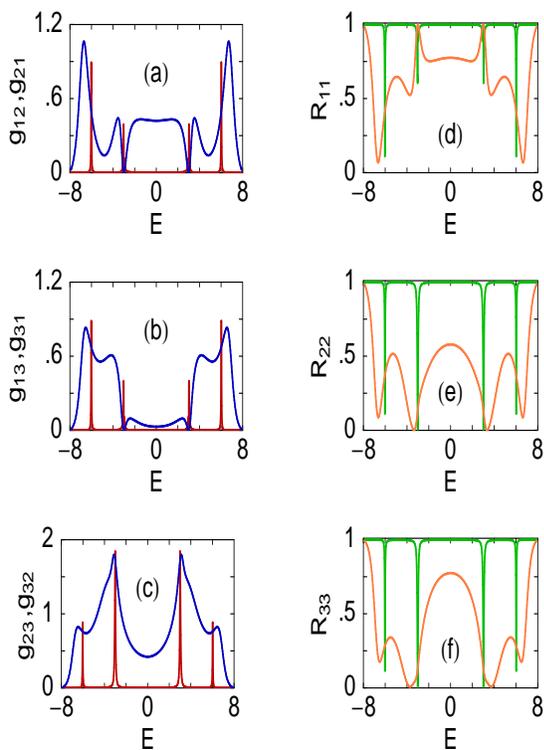}}\par}
\caption{(Color online). Three-terminal conductance $g_{pq}$ and 
reflection probability $R_{pp}$ as a function of the energy $E$ of the 
benzene molecule with $\phi=0$. The red and green curves correspond to 
the results for the weak-coupling limit, while the blue and reddish yellow 
lines represent the results for the limit of strong coupling.}
\label{threecond}
\end{figure}
As representative examples, in Fig.~\ref{threecond} we present the 
results, where the first column gives the variation of conductance 
$g_{pq}$ and the second column represents the nature of reflection 
probability $R_{pp}$. All these results are computed for $\phi=0$.
From the conductance spectra it is observed that the conductances
exhibit fine resonant peaks (red curves) for some particular energies 
in the limit of weak-coupling, while they get broadened (blue curves) 
as long as the coupling strength is enhanced to the strong-coupling
limit. The explanation for the broadening of the resonant peaks is
exactly similar as described earlier in the case of two-terminal
molecular system. A similar effect of molecular coupling to the side
attached leads is also noticed in the variation of reflection 
probability versus the energy spectra (right column of 
Fig.~\ref{threecond}). Since in this three-terminal molecular system
the leads are connected asymmetrically to the molecule i.e., the path
length between the leads are different from each other, all 
the conductance spectra are different in nature. It is also observed
that the heights of the different conductance peaks are not identical
and they get reduced significantly compared to the two-terminal case.
This is solely due to the effect of quantum interference among the 
different arms of the molecular ring. Now, in the variation of 
reflection probabilities, we also get the complex structure like as 
the conductance spectra. For this three-terminal system since the 
reflection probability is not related to the transmission probability 
simply as in the case of a two-terminal system, it is not necessarily 
true that $R_{pp}$ shows picks or dips where $g_{pq}$ has dips or picks.
It depends on the combined effect of $T_{pq}$'s.

\subsubsection{Effects of magnetic flux $\phi$}

In order to describe the dependence of magnetic flux on conductances
\begin{figure}[ht]
{\centering \resizebox*{8cm}{10.7cm}
{\includegraphics{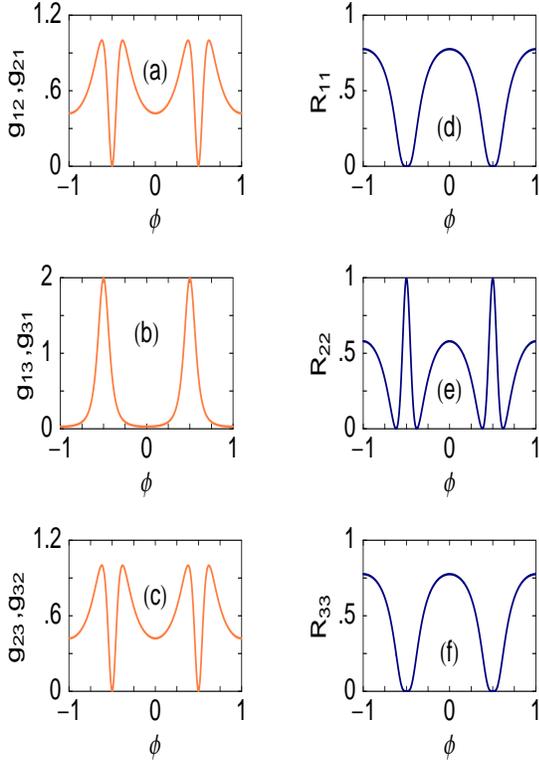}}\par}
\caption{(Color online). Three-terminal conductance $g_{pq}$ (reddish 
yellow) and reflection probability $R_{pp}$ (blue) in the limit of 
strong-coupling as a function of $\phi$ of the benzene molecule with 
$E=0$.}
\label{threecondphi}
\end{figure}
and reflection probabilities, in Fig.~\ref{threecondphi} we plot the
results for the three-terminal molecular system in the limit of
strong-coupling. The first column corresponds to the conductance
($g_{pq}$), while the second column represents the results for the
reflection probability ($R_{pp}$). All the results are done for the
typical energy $E=0$ and they show different complex spectra. The 
conductances and reflection probabilities
vary periodically exhibiting $\phi_0$ ($=1$) flux-quantum periodicity.
From our results we see that, at $\phi=\phi_0/2$, the conductances 
$g_{12}$ ($g_{21}$) and $g_{23}$ ($g_{32}$) drop exactly to zero i.e., 
$T_{12}= T_{23}=0$. While, $g_{13}$ ($g_{31}$) gets the value $2$ for 
this typical value of $\phi$. This vanishing transmission probability 
at $\phi=\phi_0/2$ will not always appear for the other choices of 
the energy $E$ in our asymmetrically connected three-terminal molecular
system. On the other hand, here it is important to note that, for our 
symmetrically connected two-terminal molecular system the transmission 
probability always vanishes for the flux $\phi=\phi_0/2$, since
for this typical value of $\phi$, a $\pi$ phase shift takes place 
among the waves traversing through the upper and lower arms of the 
molecular ring which provides zero transmission amplitude.

\subsubsection{Current-voltage characteristics: Transistor operation}

Finally, we describe the current-voltage characteristics for this
three-terminal molecular system and try to illustrate how it can be 
operated as a transistor.

The current $I_p$ passing through any lead-$p$ is obtained by 
integration procedure of the transmission function $T_{pq}$ (see 
Eq.~(\ref{equ10})), where the individual contributions from  the 
other two leads have to be taken into account. 
\begin{figure}[ht]
{\centering \resizebox*{7.5cm}{4.7cm}
{\includegraphics{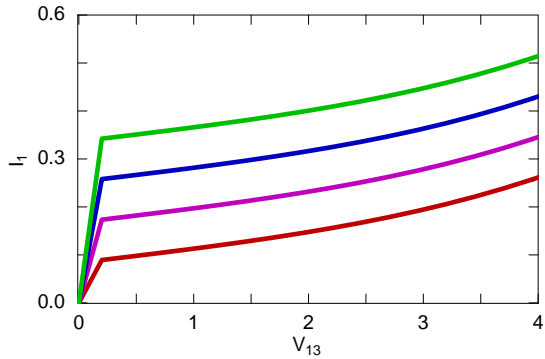}}\par}
\caption{(Color online). Current $I_1$ as a function of $V_{13}$ 
($=V_1-V_3$) for constant $V_{12}$ ($=V_1-V_2$) for the three-terminal 
molecular system in the limit of strong-coupling with $\phi=0$. The 
red, magenta, blue and green curves correspond to $V_{12}=0.2$, $0.4$, 
$0.6$ and $0.8$, respectively.}
\label{threecurr1}
\end{figure}
To be more precise, we can write the current expression for the 
three-terminal molecular device where one of the terminals serves as 
a voltage as well as a current probe~\cite{datta} in the form 
$I_p=\sum \limits_q g_{pq} \left(V_p-V_q\right) \equiv \sum 
\limits_q g_{pq} V_{pq}$, where $V_{pq}=\left(V_p-V_q\right)$ 
is the voltage difference between the lead-$p$ and lead-$q$.

In Fig.~\ref{threecurr1}, we plot the current $I_1$ in the lead-$1$ as
a function of $V_{13}$ for constant $V_{12}$ in the limit of strong
molecular coupling considering $\phi=0$. The red, magenta, blue and
green curves correspond to the currents for $V_{12}=0.2$, $0.4$, $0.6$
and $0.6$, respectively. From the results it is observed that, for 
a constant voltage difference between the lead-$1$ and lead-$2$, the
current $I_1$ initially rises to a large value when $V_{13}$ starts
to increase from zero value, and after that, it ($I_1$) increases very 
slowly with the rise of $V_{13}$ and eventually saturates. 
\begin{figure}[ht]
{\centering \resizebox*{7.5cm}{4.7cm}
{\includegraphics{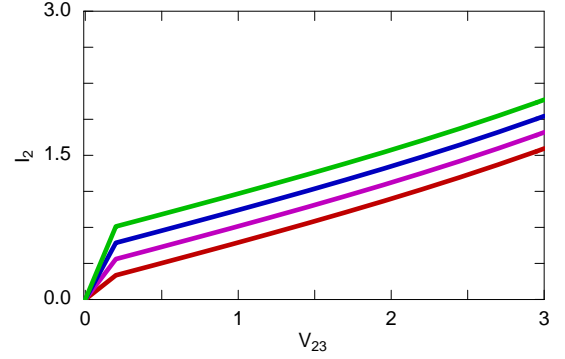}}\par}
\caption{(Color online). Current $I_2$ as a function of $V_{23}$ 
($=V_2-V_3$) for constant $V_{12}$ for the three-terminal molecular 
system in the limit of strong-coupling with $\phi=0$. The red, magenta, 
blue and green curves correspond to $V_{12}=0.4$, $0.8$, $1.2$ and 
$1.6$, respectively.}
\label{threecurr2}
\end{figure}
On the other hand, for a constant lead-$1$ to lead-$3$ voltage difference,
the current $I_1$ increases gradually as we increase $V_{12}$, which is 
clearly described from the four different curves in Fig.~\ref{threecurr1}.
Quite similar behavior is also observed in the variation of the current
\begin{figure}[ht]
{\centering \resizebox*{7.5cm}{4.7cm}
{\includegraphics{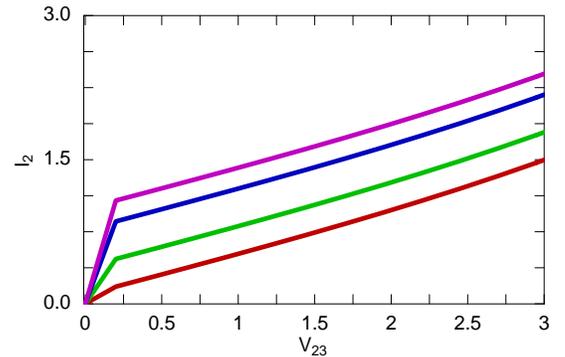}}\par}
\caption{(Color online). Current $I_2$ as a function of $V_{23}$ for 
constant $I_1$ for the three-terminal molecular system in the limit of 
strong-coupling with $\phi=0$. The red, green, blue and magenta curves 
correspond to $I_1=0.09$, $0.36$, $0.73$ and $0.92$, respectively.}
\label{threecurr3}
\end{figure}
$I_2$ as a function of $V_{23}$ for constant $V_{12}$. The results are
shown in Fig.~\ref{threecurr2}, where the currents are calculated for
the strong-coupling limit in the absence of any magnetic flux $\phi$.
The red, magenta, blue and green lines represent the currents for 
$V_{12}=0.4$, $0.8$, $1.2$ and $1.6$, respectively. Comparing the
results plotted in Figs.~\ref{threecurr1} and \ref{threecurr2}, it
is clearly observed that the current in the lead-$2$ is much higher
than the current available in the lead-$1$ for the entire voltage
range. This is solely due to the effect of quantum interference 
among the electronic waves passing through different arms of the
molecular ring, and, we can manifest that for a fixed molecular
coupling, the current amplitude significantly depends on the positions
of the different leads.

At the end, we illustrate the results plotted in Fig.~\ref{threecurr3},
where the variation of the current $I_2$ is shown as a function of
$V_{23}$ for the constant current $I_1$. The currents ($I_2$) are 
calculated for the strong-coupling limit considering $\phi=0$, where
the red, green, blue and magenta curves correspond to fixed $I_1=0.09$,
$0.36$, $0.73$ and $0.92$, respectively. For a constant $V_{23}$, 
current through the lead-$2$ increases gradually as we increase the
current $I_1$ which is clearly visible from the four different curves
in this figure. These current-voltage characteristics are quite analogous 
to a macroscopic transistor. Thus, in short, we can predict that this 
three-terminal molecular system can be operated as a transistor and 
we may call it a molecular transistor. Like a conventional macroscopic 
transistor, the three different terminals of the molecular transistor 
can be treated as emitter, base and collector. Here, the important point 
is that, since all these three terminals are equivalent to each other, 
any one of them can be considered as an emitter or base or collector.
Not only that, for this molecular transistor as there is only one type
of charge carrier, which is electron, the conventional biasing method
is not required. These features provide several key ideas which motivate
us to develop a molecular transistor rather than the traditional one.

All the above current-voltage characteristics for the three-terminal 
quantum system are studied only for the limit of strong molecular 
coupling. Exactly similar features, except the current amplitude, 
are also observed for the case of weak-coupling limit, and in the 
obvious reason here we do not plot these results once again.  

\section{Concluding remarks}

To summarize, we have explored multi-terminal electron transport through
a benzene molecule threaded by a magnetic flux $\phi$. The molecular 
system is described by a simple tight-binding Hamiltonian and all the 
calculations are done through the Green's function approach. We have 
numerically calculated the conductances, reflection probabilities and 
current-voltage characteristics as functions of the molecular coupling 
strength and magnetic flux $\phi$. Following a brief description of 
electron transport in two-terminal molecular system, we have illustrated 
in detail the behavior of electron transport in three-terminal molecular 
system. Very interestingly we have seen that the three-terminal 
benzene molecule can be operated as an electronic transistor, and we
call it as a molecular transistor. These three terminals are analogous 
to the emitter, base and collector as defined in traditional transistor. 
All these features of electron transport may be utilized in fabricating 
nano-electronic devices and our detailed investigation can provide a 
basic theoretical framework to characterize electron transport in any 
complicated multi-terminal quantum system.

Instead of a benzene molecule if we consider a small mesoscopic ring
and attach it asymmetrically to three different terminals then it can
also be operated as a transistor. Only the results presented here 
change numerically with the ring size, but all the basic features 
remain exactly invariant. To be more specific, it is important to 
note that, in real situation the experimentally achievable rings 
have typical diameters within the range $0.4$-$0.6$ $\mu$m. In such 
a small ring, unrealistically very high magnetic fields are required 
to produce a quantum flux. To overcome this situation, Hod {\em et al.} 
have studied extensively and proposed how to construct nanometer scale 
devices, based on Aharonov-Bohm interferometry, those can be operated 
in moderate magnetic fields~\cite{baer4,baer5,baer6,baer7}.

In the present paper we have done all the calculations by ignoring
the effects of the temperature, electron-electron correlation, etc. 
Due to these factors, any scattering process that appears in the
molecular ring would have influence on electronic phases, and, in
consequences can disturb the quantum interference effects. Here we
have assumed that, in our sample all these effects are too small, and
accordingly, we have neglected all these factors in this particular 
study.

The importance of this article is mainly concerned with (i) the simplicity 
of the geometry and (ii) the smallness of the size. 

\vskip 0.3in
\noindent
{\bf\small ACKNOWLEDGMENT}
\vskip 0.2in
\noindent
I acknowledge with deep sense of gratitude the illuminating comments 
and suggestions I have received from Prof. Shreekantha Sil during the 
calculations.

\end{document}